\documentclass[12pt,letter]{article}

\usepackage{graphicx, epsfig, color}
\def\eslt{E_T^{\rm miss}}
\def\to{\rightarrow}

\def\bi{\begin{itemize}}
\def\ei{\end{itemize}}
\def\te{\tilde e}

\def\tu{\tilde u}

\def\tb{\tilde b}

\def\tst{\tilde t}
\def\ttb{t\bar{t}}

\def\tg{\tilde g}

\def\tq{\tilde q}
\def\tw{\widetilde W}
\def\tz{\widetilde Z}

\def\be{\begin{equation}}  
\def\ee{\end{equation}}  
\def\bea{\begin{eqnarray}}  
\def\eea{\end{eqnarray}}  
\def\beas{\begin{eqnarray*}}  
\def\eeas{\end{eqnarray*}}  
\newcommand\prd[3]{{\it Phys.\ Rev.\ }{\bf D #1} (#2) #3}
\newcommand\prep[3]{{\it Phys.\ Rept.\ }{\bf #1} (#2) #3}

\newcommand\plb[3]{{\it Phys.\ Lett.\ }{\bf B #1} (#2) #3}
\newcommand\jhep[3]{{\it J. High Energy Phys.\ }{\bf #1} (#2) #3}
\newcommand\app[3]{{\it Astropart.\ Phys.\ }{\bf #1} (#2) #3}

\newcommand\npb[3]{{\it Nucl.\ Phys.\ }{\bf B #1} (#2) #3}

\newcommand\arnps[3]{{\it Ann.\ Rev.\ Nucl.\ Part.\ Sci.}{\bf  #1} (#2) #3}
\newcommand\ppnp[3]{{\it Prog.\ Part.\ Nucl.\ Phys.}{\bf  #1} (#2) #3}
\newcommand\jpg[3]{{\it J.\ Phys.\ {\bf G}}{\bf  #1} (#2) #3}
\newcommand{\hepph}[1]{hep-ph/#1}

\begin{document}
\begin{titlepage}
\begin{flushright}
FSU-HEP-080530\\
NSF-KITP-08-96\\
MADPH-08-0514
\end{flushright}

\vspace{0.5cm}
\begin{center}
{\Large \bf 
SUSY backgrounds to Standard Model\\
calibration processes at the LHC
}\\ 
\vspace{1.2cm} \renewcommand{\thefootnote}{\fnsymbol{footnote}}
{\large Howard Baer$^{1,3}$\footnote[1]{Email: baer@hep.fsu.edu },
Vernon Barger$^{2,3}$\footnote[2]{Email: barger@pheno.wisc.edu}, 
Gabe Shaughnessy$^2$\footnote[3]{Email: gshau@hep.wisc.edu}} \\
\vspace{1.2cm} \renewcommand{\thefootnote}{\arabic{footnote}}
{\it 
1. Dept. of Physics,
Florida State University, Tallahassee, FL 32306, USA \\
2. Dept. of Physics,
University of Wisconsin, Madison, WI 53706, USA \\
3. Kavli Institute for Theoretical Physics, University of California, Santa Barbara, CA 93106, USA\\

}

\end{center}

\vspace{0.5cm}
\begin{abstract}
\noindent 
One of the first orders of business for LHC experiments after
beam turn-on will be to calibrate the detectors using well understood
Standard Model (SM) processes such as $W$ and $Z$ production 
and $\ttb$ production. 
These familiar SM processes can be used to calibrate the electromagnetic
and hadronic calorimeters, and also to calibrate the associated missing transverse
energy signal.
However, the presence of new physics may already 
affect the results coming from these standard benchmark processes.
We show that the presence of relatively low mass supersymmetry (SUSY)
particles may give rise to significant
deviations from SM predictions of $Z+$jets and $W+$jets events for
jet multiplicity $\ge 4$ or $\ge 5$, respectively. Furthermore, the presence
of low mass SUSY may cause non-standard deviations to appear
in top quark invariant and transverse mass distributions.
Thus, effects that might be construed as detector mal-performance 
could in fact be the presence of new physics.
We advocate several methods to check when new physics might be 
present within SM calibration data.
\vspace*{0.8cm}


\end{abstract}


\end{titlepage}

\section{Introduction}
\label{sec:intro}

The CERN Large Hadron Collider (LHC) is expected soon to begin circulation of beams,
with proton-proton collisions at center-of-mass energy $\sqrt{s}\sim 10$ TeV, followed 
by a physics run at $\sqrt{s}=14$ TeV. This energy scale ought to 
be sufficient to explore the mechanism behind the breakdown of electroweak symmetry (EWSB)\cite{lhcreview}.
Expectations are high that the upcoming LHC era will provide the necessary data
to construct a new paradigm for the laws of physics as we know them, and new physics
beyond that of the Standard Model (SM) is widely anticipated.
Numerous theoretical constructs have been created, including weak scale supersymmetry\cite{wss},
theories with extra dimensions\cite{hp}, little Higgs models\cite{lh}, {\it etc.}\cite{lykken}.
Which- if any- of these theories correctly describes nature at the TeV scale will be determined by LHC  experiments~\cite{Hubisz:2008gg}.

However, before new physics searches can commence at the LHC, it will be necessary to fully understand the
responses of the ATLAS and CMS detectors to particle physics scattering events.
Towards this end, the experimental groups will first focus their interest on well-understood
SM processes. These will include the following:
\begin{itemize}

\item Detect familiar di-jet or multi-jet events and compare the
measured jet $E_T$, $\eta $(=pseudorapidity) and invariant mass distributions to
expectations from QCD theory and Monte Carlo event generators.

\item Detect the $Z\to\ell\bar{\ell}$ events ($\ell =e$ or $\mu$ initially) 
and verifying that the $Z$-peak appears with the correct mass and width as measured by
LEP/Tevatron experiments. One also wants to study the associated multi-jet
activity and use the rates as a tune for Monte Carlo generators. In particular, 
$Z\to\ell\bar{\ell}+$jets production should serve as an excellent calibrator for
$Z\to \nu\bar{\nu}+$jets production, which is one of the most important backgrounds
for SUSY jets $+\eslt$ searches.

\item Detect $W\to\ell\nu_\ell$ events,
verify the expected transverse mass distribution and investigate again the
associated multiplicity of jets expected.

\item Detect $t\bar{t}$ production and check that the LHC-measured $m_t$ 
value is in accord with the most recent Tevatron results. 
The simplest channel to begin with is
$t\to b\ell\nu_{\ell}$ while the other $t$ decays hadronically, thus allowing 
a direct $m_t$ reconstruction with relatively few ambiguities. 
\end{itemize} 

Once these familiar processes are measured, and it is ascertained that expected results are
in accord with previous LEP/Tevatron measurements and SM theory, then the hunt for {\it deviations}
from the SM can be made and the search for new physics will have begun.
In addition, if a high degree of confidence is gained in the observation of 
$W$ and $Z$ production, then these reactions may serve as a {\it luminosity monitor}.

The rate for $W^\pm$ boson production at the LHC is $\sim 100$ nb,
while $Z$ production occurs at the $\sim 40$ nb rate.
The $pp\to \ttb X$ cross section for $m_t=175$ GeV is $\sim 800$ pb. 
Thus, with only $0.01$ fb$^{-1}$ of
integrated luminosity, sizable $W$, $Z$ and $\ttb$ signals are expected.
 
In this paper, we wish to make the point that it is possible that {\it new physics}
signals at the LHC may be so large that they may already make their presence felt in these
baseline SM reactions. In this case, they would potentially disrupt efforts
at calibrating both detectors and Monte Carlo event generators to expected SM signal levels. 

As an example, we consider a relatively low mass spectrum of 
supersymmetric particles. Weak scale supersymmetry is one the main classes
of new physics signals to be searched for at the LHC. We adopt several 
simple SUSY case studies and show that these may cause distortions in the 
$Z+$jets, $W+$jets and $\ttb$ signal channels.

In our simulations we adopt the event generator Isajet 7.76\cite{isajet} to model $Z+$jets, 
$W+$jets, $\ttb$ and weak boson pair production at the LHC, along with superparticle production.
By prescribing a non-zero $q_T$ distribution for $W$ and $Z$ production, 
Isajet calculates $W$ or $Z$ plus one parton emission using
exact QCD matrix elements. Further jets are developed via the parton shower algorithm.
This procedure generates the expected high $q_T(W)$ and $q_T(Z)$ distributions.
The multi-jet production via this method agrees rather well with 
overall rates for weak boson plus multi-jet production at the Tevatron and
is in qualitative accord with the merged NLO weak boson production plus
parton shower approach advocated in Ref. \cite{br}.  

We use Isajet also for the simulation of signal and 
background events at the LHC. A toy detector simulation is employed with
calorimeter cell size
$\Delta\eta\times\Delta\phi=0.05\times 0.05$ and $-5<\eta<5$. The hadronic calorimeter
energy resolution is taken to be $80\%/\sqrt{E}+3\%$ for $|\eta|<2.6$ and
$100\%/\sqrt{E}+5\%$ for the forward region with $|\eta|>2.6$. 
The electromagnetic calorimeter energy resolution
is assumed to be $3\%/\sqrt{E}+0.5\%$. We use a UA1-like jet finding algorithm
with jet cone size $R=0.4$ and require that $E_T(jet)>50$ GeV and
$|\eta (jet)|<3.0$. Leptons are considered
isolated if they have $p_T(e\ or\ \mu)>20$ GeV and $|\eta|<2.5$ with 
visible activity within a cone of $\Delta R<0.2$ of
$\Sigma E_T^{cells}<5$ GeV. The strict lepton isolation criterion helps reduce
multi-lepton backgrounds from $c\bar c$ and $b\bar{b}$ production.

We identify a hadronic cluster with $E_T>50$ GeV and $|\eta(j)|<1.5$
as a $b$-jet if it contains a $B$ hadron with $p_T(B)>15$ GeV and
$|\eta (B)|<3$ within a cone of $\Delta R<0.5$ about the jet axis. We
adopt a $b$-jet tagging efficiency of 60\% and assume that
light quark and gluon jets may be mis-tagged as $b$-jets with a
probability $1/150$ for $E_T<100$ GeV, $1/50$ for $E_T>250$ GeV, 
with a linear interpolation for $100$ GeV$<E_T<$ 250 GeV\cite{xt}. 

We will compare expectations for SM calibration reactions against
a case study from the well-known minimal supergravity
(mSUGRA or CMSSM) model, plus two other case studies with 
non-universal soft SUSY breaking terms. In mSUGRA, given a set of parameters
\be
m_0,\ m_{1/2},\ A_0,\ \tan\beta ,\ sign(\mu ) ,
\ee
one may calculate the entire superparticle mass spectrum, along with
sparticle branching fractions and production cross sections.
The Isajet program is used to generate the associated LHC collider
events. In our work we adopt three case studies listed in Table
\ref{tab:susy}. The mSUGRA point P1 is chosen to give a 
sparticle mass spectrum which yields a modest rate for production
of real $Z$ bosons from sparticle cascade decays.
Point P2 is similar to point P1 except that it is within the context of
non-universal higgs models (NUHM)\cite{nuhm}, wherein the superpotential $\mu$ 
parameter is fixed to a higher value (400 GeV) so that gluino and squark decays to higher
chargino and neutralino states (which in turn may decay to real $Z$ bosons)
are kinematically suppressed. Thus, point P1 is expected to yield a 
significant rate for real $Z$ bosons in SUSY events, while point P2 is constructed
so that $Z$ production in cascade decays is suppressed.
Finally, point P3 is engineered so as to provide a rather high rate of 
$Z$ boson production in cascade decays. P3 has GUT scale gaugino masses
$M_2=M_3=200$ GeV, while the $U(1)_Y$ gaugino mass $M_1$ is just 120 GeV.
The non-universal gaugino masses allow $\tz_2\to \tz_1 Z$ decay, and since $\tz_2$
is produced at a high rate in gluino and squark cascade decays, real $Z$ bosons are
expected to be very common in LHC events for point P3.
%
\begin{table}
\begin{center}
\begin{tabular}{lccc}
\hline
parameter & P1 & P2 & P3\\
\hline
$m_0$     & 150 & 150 & 200\\
$m_{1/2}$ & 170 & 170 & 200 \\
$M_1$ & 170 & 170 & 120 \\
$A_0$ & -200 & -200 & 0 \\
$\tan\beta$ & 20 & 20 & 20 \\
$\mu$ & 262.4 & 400 & 261.5 \\
$m_{\tg}$   & 432.0 & 430.3 & 503.8 \\
$m_{\tu_L}$ & 416.7 & 417.1 & 493.4 \\
$m_{\tst_1}$& 256.9 & 302.0 & 352.2 \\
$m_{\tb_1}$ & 364.6 & 373.9 & 443.4 \\
$m_{\te_L}$ & 195.1 & 192.6 & 244.1 \\
$m_{\tw_2}$ & 290.8 & 415.2 & 294.0 \\
$m_{\tw_1}$ & 117.4 & 126.3 & 137.6 \\
$m_{\tz_4}$ & 290.9 & 414.7 & 293.3 \\ 
$m_{\tz_3}$ & 271.1 & 404.2 & 270.8 \\ 
$m_{\tz_2}$ & 117.7 & 126.2 & 137.3 \\ 
$m_{\tz_1}$ &  64.4 & 66.3  & 44.8 \\ 
$m_A$       & 293.4 & 293.4 & 327.1 \\
$m_h$       & 108.2 & 108.3 & 107.7 \\ 
\hline
$\sigma\ [{\rm fb}]$ & $2.2\times 10^5$ & $2.2\times 10^5$ & $1.0\times 10^5$ \\
\hline
\end{tabular}
\caption{Masses and parameters in~GeV units
for three cases studies points P1, P2 and P3
using Isajet 7.76 with $m_t=171.0$ GeV and $\mu >0$. 
We also list the 
total tree level sparticle production cross section 
in fb at the LHC.
}
\label{tab:susy}
\end{center}
\end{table}

\section{$Z+$jets production at LHC}
\label{sec:zj}

To select $Z$-boson candidate events, we require the following 
\textbf{\boldmath$Z$ Cuts:}
\bi
\item 2 isolated opposite-sign, same-flavor (OSSF) leptons,
\item $80\ {\rm GeV}< m(\ell\bar{\ell})<100\ {\rm GeV}$,
\item  $p_T(j_1)> 100\ {\rm GeV}$,
\item  $p_T(j_2,...,j_n)> 50\ {\rm GeV}$,
\ei
where the $n$ jets in each event, $j_{1},j_{2},...,j_{n}$, are $p_{T}$ ordered.

The staggered jet cuts require the first hard parton emission to be at the
highest $p_T$, which is modeled by the exact QCD $Z+parton$ matrix element.
Further jets at lower $p_T>50$ GeV are generated by the parton shower (PS) algorithm.

Once the electromagnetic calorimeter and muon chambers are calibrated so that the 
$Z\to e^+e^-$ and $Z\to\mu^+\mu^-$ peaks are in accord with previous 
LEP/Tevatron measurements, then it will be important to map out the 
jet activity associated with these events. This will help to calibrate the
hadronic calorimeter and also to tune the 
weak boson plus jets Monte Carlo programs to the data. Further, the
rate for $Z\to\nu_i\bar{\nu}_i+$jets production should occur at
six times the $Z\to e^+e^- +$jets rate, so this process gives a good estimate
of one of the most important backgrounds to the SUSY jets$+\eslt$ channel.

\begin{figure}[htbp]
\begin{center}\vspace{.5in}
\includegraphics[angle=0,width=0.59\textwidth]{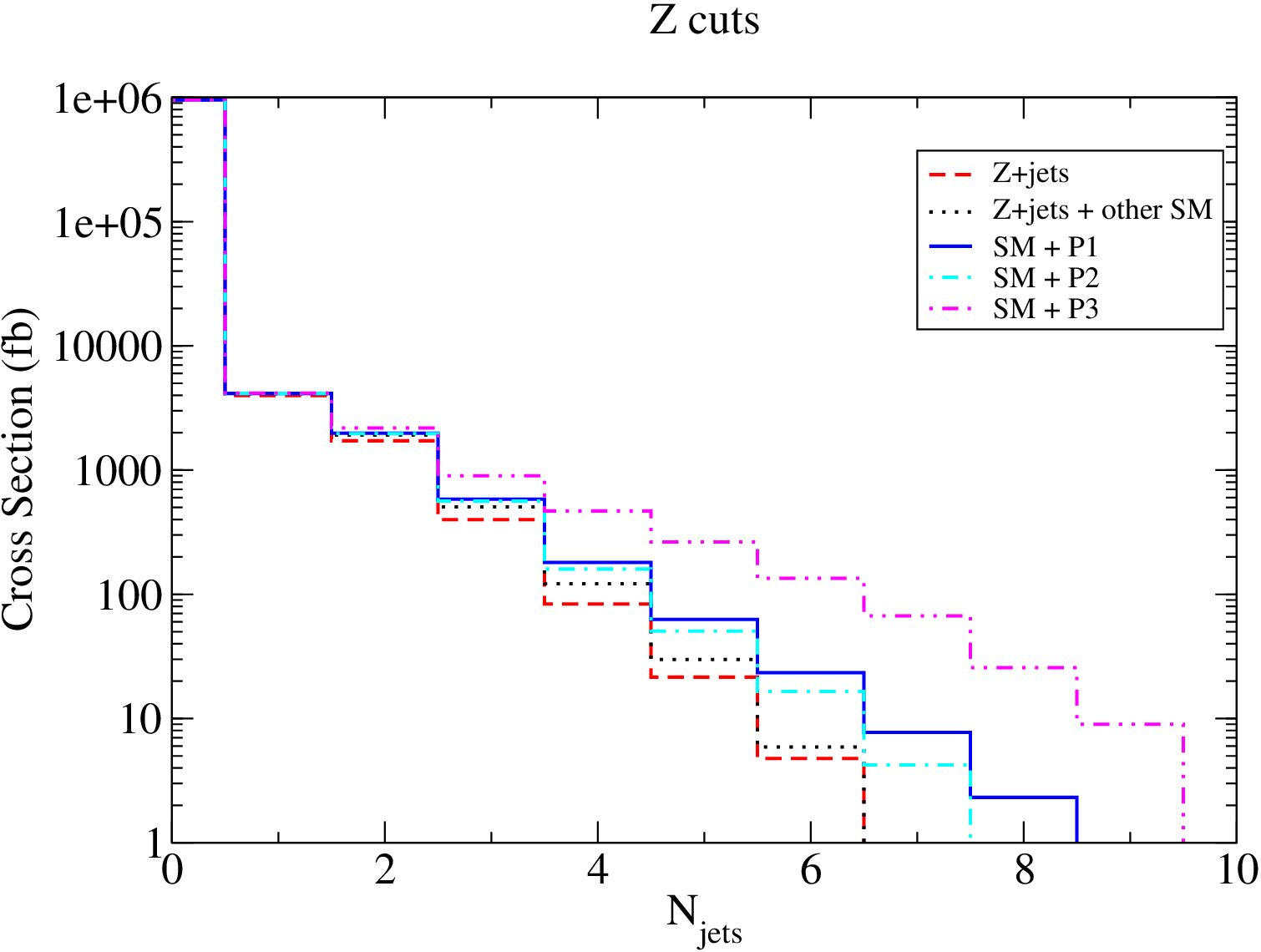}
\caption{Jet multiplicity in $Z\to\ell\bar{\ell}$ events from $Z$ production, 
$\ttb$ and $VV$ production and sparticle production from SUSY points P1, P2 and P3.
}
\label{fig:zj}
\end{center}
\end{figure}

In Fig. \ref{fig:zj} we plot the expected jet multiplicity coming from
$Z\to \ell^+\ell^- +$jets events after $Z$ cuts. The red-dashed histogram gives the
Isajet prediction for the jet multiplicity from $Z+$jets production alone, while the
black-dashed histogram gives the $Z$ {\it plus} $t\bar{t}$ and $VV$ contributions
(here, $V=W$ or $Z$). 
The blue-solid histogram shows the SM contribution plus SUSY from point P1.
In point P1, the branching fraction of $\tz_3\to\tz_1 Z$ is 11\%, 
$\tz_3\to\tz_2 Z$ is 17\% and $\tw_2\to \tw_1 Z$ is at 21\% , while
$\tz_4\to \tz_i Z$ ($i=1,2$) is at 5.7\%. 
However, the squark branching fractions to $\tz_3$, $\tz_4$ and $\tw_2$ occur
only at the 5-10\% level, while gluino branchings into these heavier -ino 
states is less than 1\%.
Thus, we expect a modest rate
for $Z+$jets production in squark cascade decays to $\tz_3$, $\tz_4$
and $\tw_2$\cite{susyz} for point P1. Point P2 has non-universal Higgs masses,
and here we have raised the superpotential Higgs mass $\mu$ to 400 GeV.
This effectively cuts off any squark or gluino decays to heavier ino states, so we expect
a low rate of $Z$ production in SUSY cascade decays. Meanwhile, point P3 has a low
GUT scale value of the $U(1)_Y$ gaugino mass $M_1$, which allows $BF(\tz_2\to \tz_1 Z)$ at
98\%. Since $\tz_2$ is produced copiously in gluino and squark cascade decays, P3 is expected to yield LHC collider events that are enriched in $Z$ bosons. 

We see from Fig. \ref{fig:zj} that the low jet multiplicity events are dominated
by $Z+$jets production as expected. As $n(jets)$ increases, $\ttb$, $WZ$
and $ZZ$ contributions give elevated predictions of jet activity in the
``$Z+$jets'' events. In addition, as we step out to higher and higher jet 
multiplicity, the SUSY contribution from point P1 becomes ever more important:
by $n(jets)=4$, the SUSY rate is comparable to direct $Z+$jets production, and
by $n(jets)=5$, SUSY of P1 dominates the event rate. 
Point P3 begins to dominate the $Z+$jets rate already at $n(jets)=3$.
Thus, in these new physics examples, a direct tuning of SM Monte Carlo programs for $Z+$jets production to data would result in an aberrant tuning, as the data would also include 
significant portions of both SM and new physics events.

Naively one would expect that the SUSY events could easily be
separated from SM background by requiring large $\eslt$. However, in the 
early stages of LHC running when detectors are first being calibrated, 
the $\eslt$ measurement may not be sufficiently reliable due to a variety of detector
effects such as mis-calibration, no calibration, under instrumented regions,
event overlap, beam-gas events, cosmic rays etc.\cite{bps}. In addition, 
OSSF dilepton events from $\ttb$ and $ZZ$ production will also have 
$\eslt$ present from neutrinos in the events.

\begin{figure}[htbp]
\begin{center}\vspace{.5in}
\includegraphics[angle=0,width=0.59\textwidth]{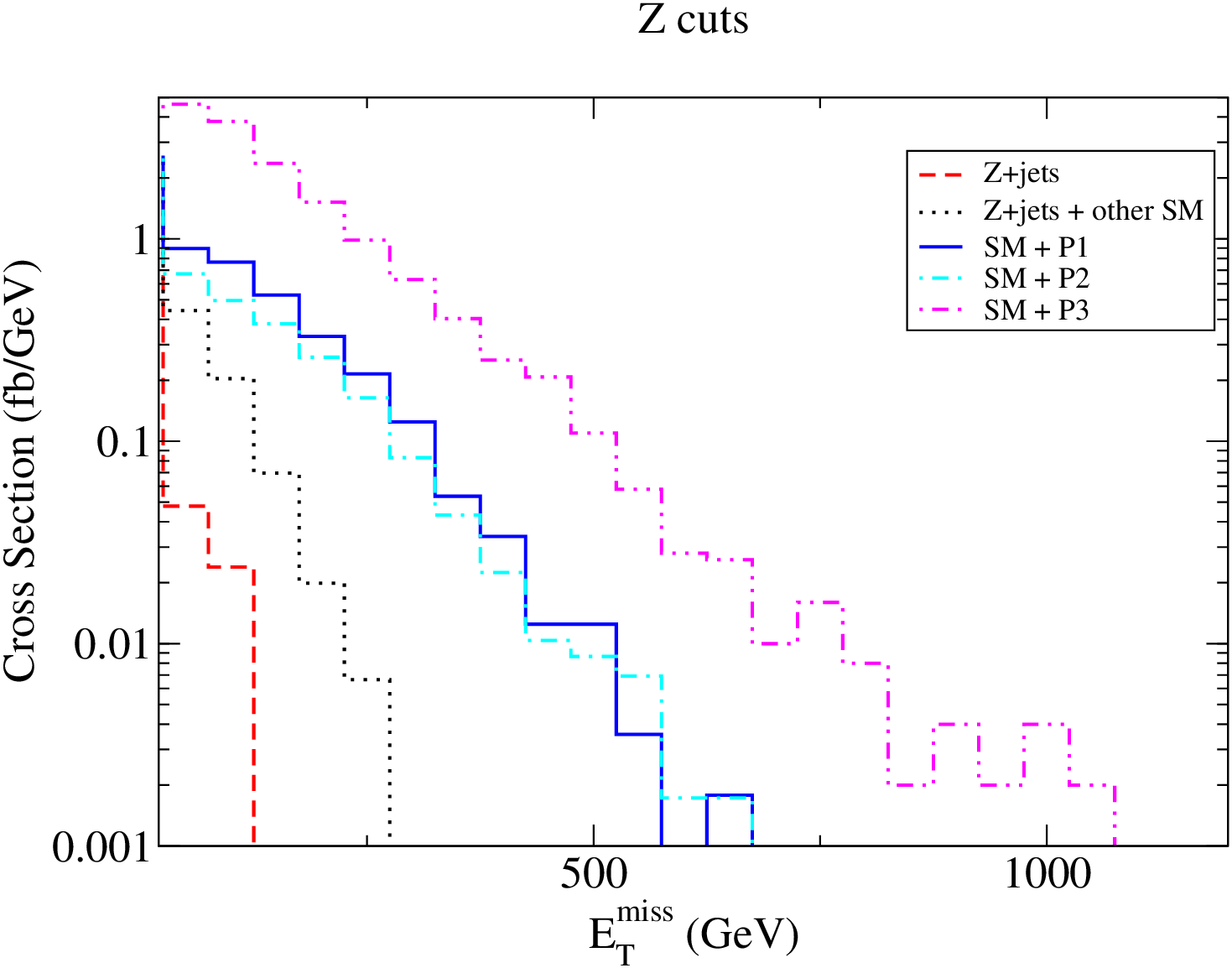}
\caption{Missing $E_T$ distribution in $Z\to\ell\bar{\ell}$ events from $Z$ production, 
$\ttb$ and $VV$ ($WW, WZ$ and $ZZ$) production and sparticle production from SUSY points 
P1, P2 and P3.
}
\label{fig:zj_etmiss}
\end{center}
\end{figure}

In Fig. \ref{fig:zj_etmiss}, we plot the expected $\eslt$ distribution
arising from $Z+$jets production (red-dashed histogram), $Z+$jets along
with $\ttb$ and $VV$ production (black-dotted histogram) and all
SM sources {\it plus} SUSY points P1, P2 and P3.
The SUSY events typically have much more than 50 GeV of $\eslt$
present. It would then be hoped that, even with initially poor $\eslt$ 
resolution, the presence of very large $\eslt$ in the $Z+$jets event sample
would alert one to the presence of new physics in this calibration process. 

\section{$W+$jets production at LHC}
\label{sec:wj}

To select $W$-boson candidate events, we require the following \textbf{\boldmath $W$ Cuts:}
\bi
\item a single isolated lepton,
\item transverse mass $40\ {\rm GeV}< m_T(\ell,\eslt )<120\ {\rm GeV}$,
\item $p_T(j_1)> 100\ {\rm GeV}$,
\item $p_T(j_2,...,j_n)> 50\ {\rm GeV}$.
\ei
Initially, it will be difficult to establish $\eslt$ as a reliable cut variable, 
so we omit it from our cuts list. However, it is used implicitly in 
constructing the lepton-missing energy transverse mass variable 
$M_T(\ell,\eslt )$. 

One of the first things to check in $W\to\ell\nu_{\ell}$ events will be that
the transverse mass distribution has the requisite shape, including a
Jacobian peak at the $W$-boson mass.  The uncertainty associated with 
the $\eslt$ measurement will lead to broadening of the distribution.
As the $\eslt$ resolution is sharpened, the $W$-boson Jacobian peak
should assume its characteristic form.
Once confidence is gained that one is really seeing $W\to\ell\nu_\ell$ 
events, then the next step will be to examine the associated jet activity.  
The $W+$jets events will be an important background to $\ttb$ searches, as
well as other new physics searches such as SUSY where one looks for
single lepton plus jets $+\eslt$ events.

\begin{figure}[htbp]
\begin{center}\vspace{.5in}
\includegraphics[angle=0,width=0.59\textwidth]{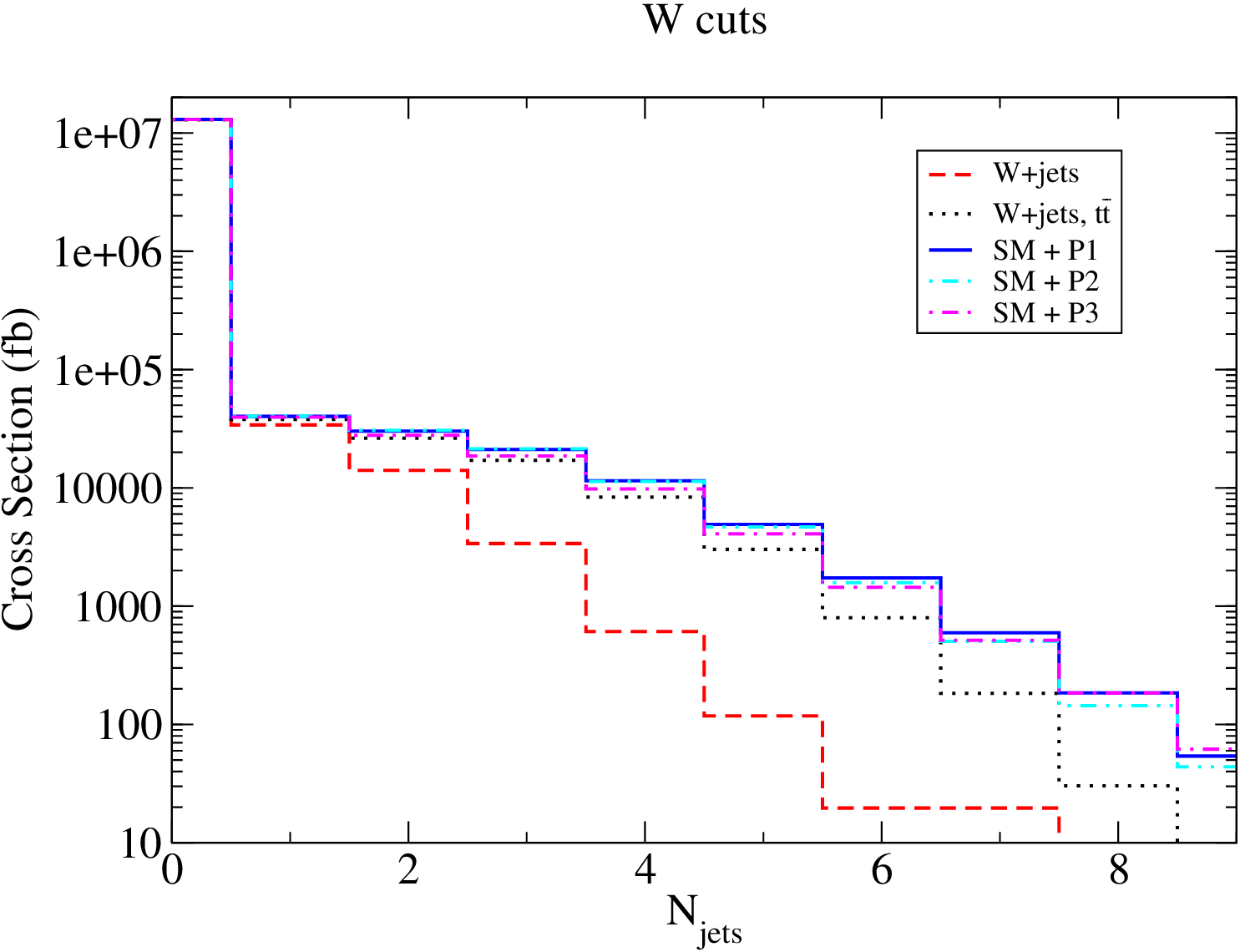}
\caption{Jet multiplicity in $W\to\ell\nu_\ell$ events from $W$ production, 
$\ttb$ and $VV$ production and sparticle production from SUSY points
P1, P2 and P3.
}
\label{fig:wj}
\end{center}
\end{figure}

We plot in Fig. \ref{fig:wj} the jet multiplicity associated with 
$W\to\ell\nu_\ell$ events after the above cuts. The red-dashed histogram
shows the contribution from SM $W+$jets (without $t\bar t$ contributions) events generated from Isajet.
The black-dotted histogram includes also the contribution from
$\ttb$ and $VV$ events. We also show histograms including $W+$jets, $\ttb$, $VV$ 
along with SUSY from points P1, P2 and P3. We see that at low $n(jets)=0,\ 1$, indeed most
events come from $W+$jets production. As we move to $n(jets)\ge 2$, the
events are increasingly dominated by $\ttb$ production. In fact, this also 
occurs at the Tevatron collider and was proposed as a search strategy for
SM top events\cite{top_njets}. The $\ttb$ contribution to $W+$jets events
is accentuated at the LHC since the $\ttb$ production rate increases by about a factor 
of $\sim 100$ from the Tevatron to the LHC, while $W$ and $Z$ production increase by
just a factor of $5$.
We also see that as $n(jets)$ moves out to very high multiplicity,
$n(jets)\ge 5$, it can be the case that SUSY makes a significant
contribution to the $n(jets)$ distribution. Thus, again, calibrating 
Monte Carlo $W$ and $\ttb$ generators to the $W+n$-jets data sample
may again lead one askew in that new physics may already be present in the
actual data event set.

\begin{figure}[htbp]
\begin{center}\vspace{.5in}
\includegraphics[angle=0,width=0.59\textwidth]{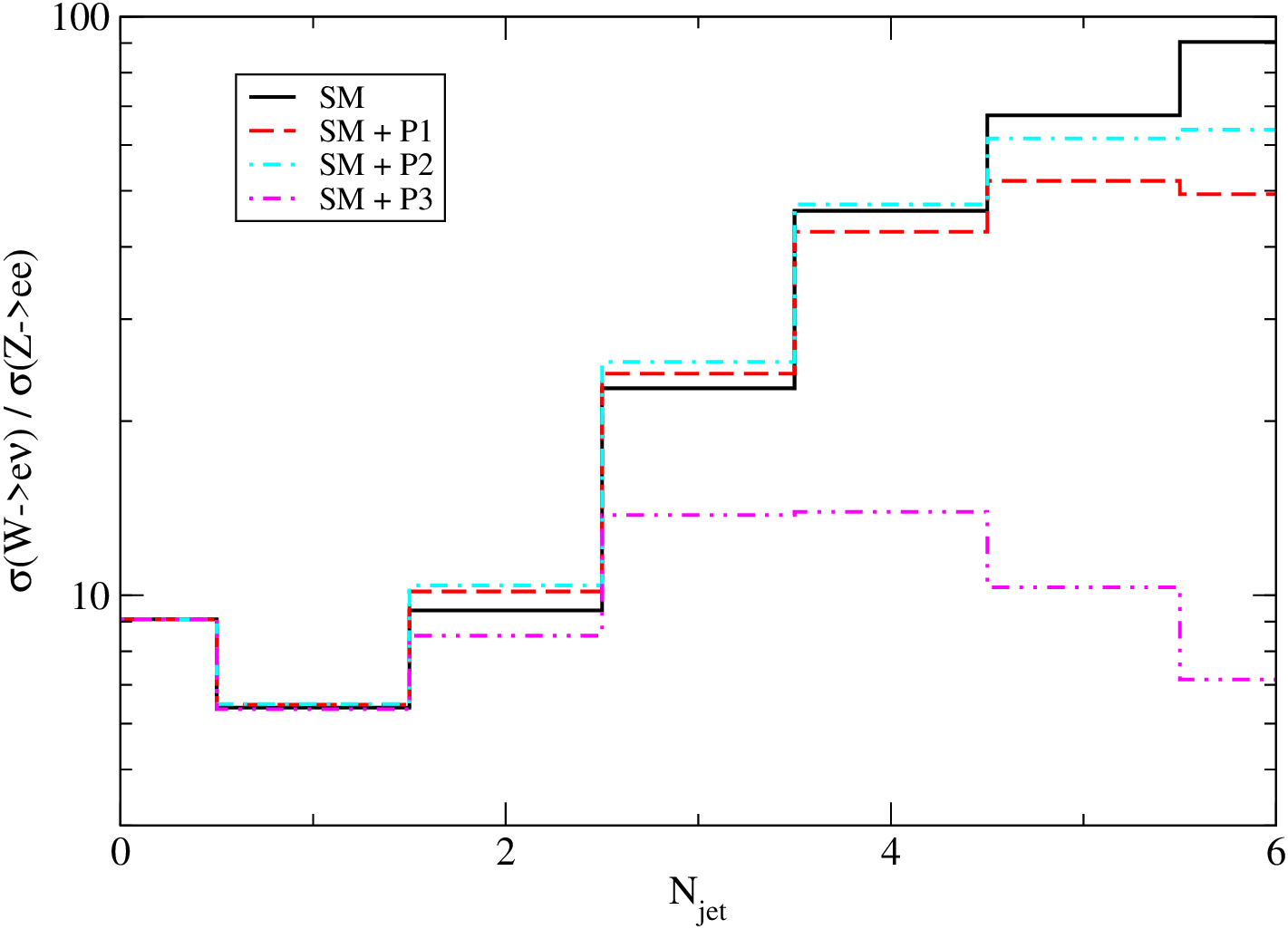}
\caption{Ratio of $W\to e\nu_e$ to $Z\to e\bar{e}$ events at the LHC versus
jet multiplicity from SM processes (black-solid histogram) and
from SM plus SUSY points P1, P2 and P3.
}
\label{fig:wz}
\end{center}
\end{figure}

One way to check early on whether the $W+$jets sample at the LHC is in accord with
SM predictions is to examine the $\sigma (W\to e\nu_e )/\sigma (Z\to e\bar{e})$ ratio
versus $n(jets)$. In the $W/Z$ ratio, the various QCD and PDF uncertainties
cancel. In Fig. \ref{fig:wz}, we show the $W/Z$ distribution
versus $n(jets)$ for the summed SM processes along with SUSY points P1, P2 and P3 
at $\sqrt{s}=14$ TeV.
From only SM events without $t\bar t$, the ratio should be at $\sigma (pp\to W)\times BF(W\to e\nu_e )/\sigma (pp\to Z)\times
BF(Z\to e\bar{e})$ is around 8.3, as evident in the first bin. This is verified in the first bin: $n(jets)=0$.
In the second bin, $n(jets)=1$, the ratio is somewhat lower due to the influence of
the different $W$-cuts versus $Z$-cuts. At $n(jets)\ge 2$, contributions from $\ttb$
production cause an increase in $W$ production relative to $Z$ production, and the ratio
steadily climbs with jet multiplicity. In the case where $Z$ production is substantial
in SUSY events, the SUSY contribution eventually will enter, and at high jet multiplicity, the
$W/Z$ ratio again starts to fall off, indicating the presence of non-SM processes in 
the calibration sample. This is especially so for the $Z$-rich events from SUSY point P3.

\section{$\ttb$ production at LHC}
\label{sec:ttb}

To select $\ttb$ candidate events, we examine the case where
$\ttb\to (b\ell\nu_\ell )+(\bar{b}q\bar{q}')$, {\it i.e.} 
one leptonic and one hadronic decay. This channel should allow for 
a simple top mass reconstruction via the invariant mass $m(jjb)$ and
the cluster transverse mass $M_T(b\ell,\eslt )$~\cite{Barger:1983jx}.

We require the following \textbf{\boldmath $\ttb\ 1\ell$ cuts:}  
\bi
\item a single isolated lepton,
\item $n(jets)\ge  4$,
\item $n(b-jets)=  1$,
\item transverse mass $40\ {\rm GeV}<  m_T(\ell,\eslt )<90\ {\rm GeV}$,
\item invariant  mass $60 \ {\rm GeV}<m(jj) <90\ {\rm GeV}$, and
\item $p_T(j_{4})> 40\ {\rm GeV}$.
\ei

The $jj$ invariant mass cut selects events where two jets reconstruct 
a hadronic $W$. To reconstruct the top mass, 
there is then an ambiguity on which $b$-jet 
is associated with the $W\to q\bar{q}'$ system. Usually we will
get this right by choosing the {\it minimum} invariant mass gained by 
combining the two jets which make up the $W$ mass with any of the remaining
jets~\cite{Barger:2006hm}.

\begin{figure}[htbp]
\begin{center}\vspace{.5in}
\includegraphics[angle=0,width=0.59\textwidth]{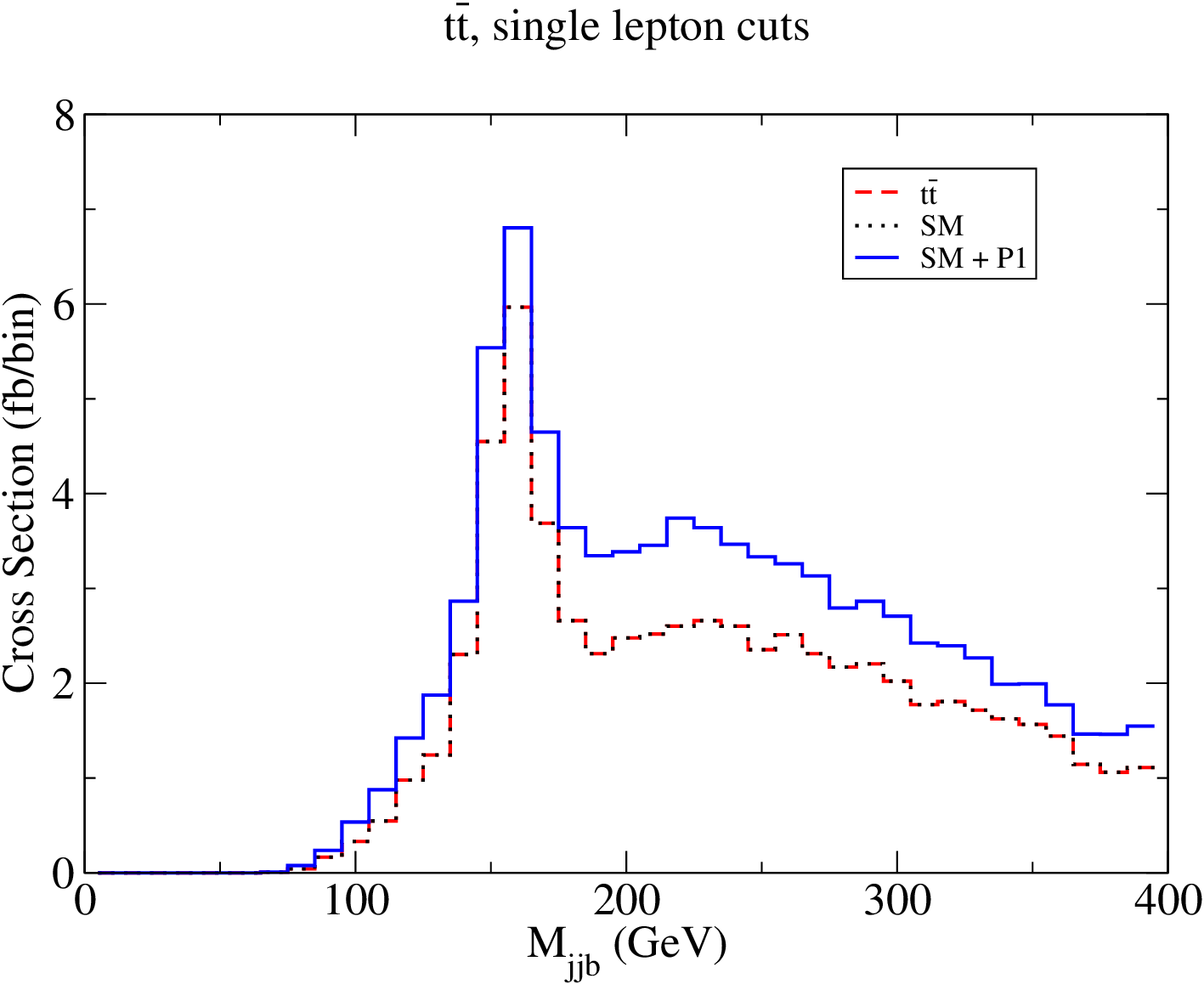}
\caption{Distribution in $min[m(bjj)]$ from $\ttb$ and $W+$jets production, 
and including SUSY production from point P2.
}
\label{fig:mjjb}
\end{center}
\end{figure}

In Fig. \ref{fig:mjjb}, we plot the invariant mass of the two jets
which combine to make up $M_W$, with the remaining jet which yields the 
minimum invariant mass.
The red-dashed histogram shows the resulting distribution from 
$\ttb$ production, and a clear peak is seen just below
the value of $m_t$. The high invariant mass tail comes from
a wrong assignment of the additional jet to the $m(jj)=M_W$ cluster
(here the correct parton is usually too soft to satisfy the jet 
finding criteria, or ends up overlapping with another jet).
We also show the $\ttb$ plus the $W+$jets contribution as the black-dotted
histogram, which essentially overlaps with the $\ttb$ plot since the
$W+$jets background is very low for this set of cuts.
We also show in Fig. \ref{fig:mjjb} the invariant mass construction
when the SUSY contributions from point P2 are added in. Here, we again see the distinct top peak,
but now the high $m(jjb)$ shoulder is much higher. In this case, the 
reconstructed top mass distribution would suffer significant shape distortion 
on the high end. A tuning of the $\ttb$ Monte Carlo to LHC top quark data
could again result in a mis-calibration.
The shape distortion can be simply characterized by the peak height to the shoulder
height. In the SM, the ratio of heights is $\sim 2$, while adding in a SUSY
contribution gives a ratio of $\sim 1.5$.

\begin{figure}[htbp]
\begin{center}\vspace{.5in}
\includegraphics[angle=0,width=0.59\textwidth]{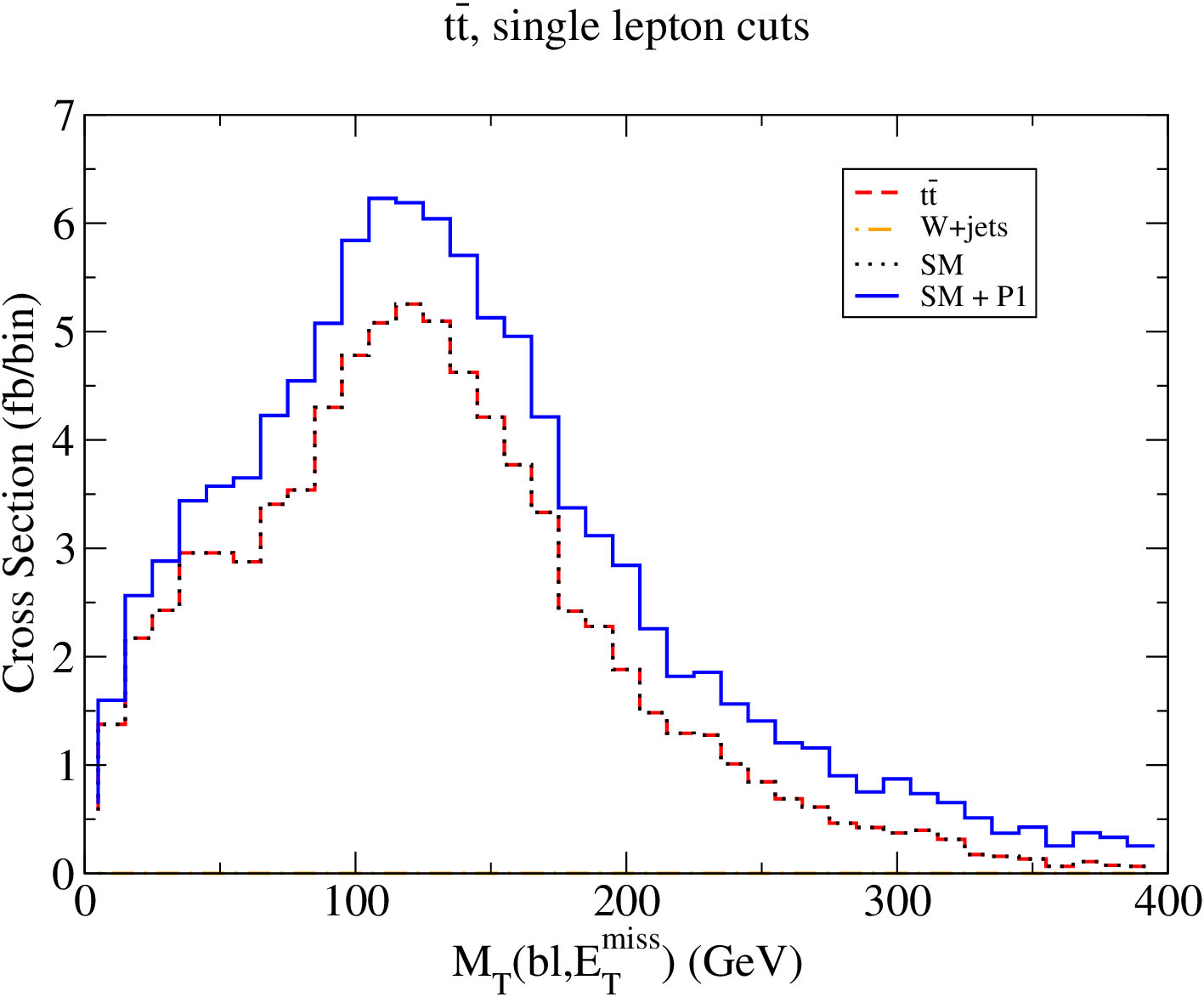}
\caption{Distribution in cluster transverse mass $m_T(b\ell,\eslt )$ from 
$\ttb$ and $W+$jets production, and from SUSY
production with
$m_0=150$ GeV, $m_{1/2}=170$ GeV, $A_0 =-200$, $\tan \beta = 20$,
$\mu >0$ and $m_t=171$ GeV.  
}
\label{fig:mcl}
\end{center}
\end{figure}

In Fig. \ref{fig:mcl} we show the cluster transverse mass
$M_T(b\ell,\eslt )$. This distribution is formed from the 
isolated lepton combined with the jet which yields the smallest invariant
mass and then constructing the cluster transverse mass with the 
$\eslt$ variable. The distribution in the ideal case would be a 
continuum bounded from above by $m_t$. 
Here, we see that the
$\ttb$ contribution has a tail extending to high
cluster transverse masses owing to mis-combination of jets with leptons, and
also from additional sources of $\eslt$, such as neutrinos and 
jet energy mis-measurement. 
The distribution
from $\ttb$ plus SUSY with point P2 has some widening associated with it, and 
so could again lead to Monte Carlo mis-tuning. 

One of the main ways to distinguish if the $\ttb$ signal is
really from $\ttb$ is to compare results from the $1\ell$ channel
to results from the dilepton channel, where both $t$ and $\bar{t}$
decay semi-leptonically. 
We will invoke the following \textbf{\boldmath $\ttb\ 2\ell$ cuts}, taken from
Ref. \cite{cms}:
\bi
\item two isolated leptons,
\item $n(jets)\ge  2$,
\item $n(b-jets)=  2$,
\item $p_T(\ell_{2})>20$ GeV, and
\item $\eslt >40$ GeV.
\ei
For now, we will not distinguish whether or not the dileptons are
opposite-sign (OS) or same-sign (SS). The ratio of single-to-dilepton
events from $\ttb$ and other SM sources should be
a very predictable quantity. However, in SUSY events, there is a high propensity
to generate states with large isolated lepton multiplicity due to the 
gluino and squark cascade decays. Thus, if there is SUSY contamination of the
$\ttb$ signal, we would expect enhanced dilepton signal rates compared to
single lepton rates. We illustrate this in Table \ref{tab:1l2l}, where the
single lepton and dilepton rates are listed in fb after $\ttb$-cuts for
single (C3 cut) and dilepton (C4 cut) events. We see a decrease in the $1\ell /2\ell$ ratios due to SUSY contamination.
%
\begin{table}
\begin{center}
\begin{tabular}{lccc}
\hline
point & $1\ell$ & $2\ell$ & $1\ell / 2\ell$ \\
\hline
SM    & 797 & 69 & 11.6 \\
SM+P1 & 1063 & 105 & 10.1 \\
SM+P2 & 983 & 106 & 9.3 \\
SM+P3 & 972 & 92 & 10.5 \\
\hline
\end{tabular}
\caption{Cross sections in fb from $t\bar t$ cuts for $1\ell$ events and $2\ell$ events from
SM sources and for SM plus contributions from mSUGRA points P1, P2 or P3, along with the single lepton to 
dilepton ratio.
}
\label{tab:1l2l}
\end{center}
\end{table}

Another possibility is to separately consider OS and SS dileptons. 
SS dileptons are expected to be rare at LHC, although they can arise from $\ttb$ 
production where $t\to b\ell^+\nu_\ell$ and $\bar{t}\to \bar{b}q\bar{q}'$, 
followed by $\bar{b}\to \bar{c}\ell^+\nu_\ell$ decay, where the lepton from
$\bar{b}$ decay is isolated, and from $B^0-\bar B^{0}$ mixing. Further sources come from secondary processes such as 
$W^\pm W^\pm$ production, which we do not simulate here.
However, in SUSY events, SS dileptons can arise from $\tg\tg$ production, where
each gluino decays to $\tw_1^+ +q\bar{q}'$ for instance\cite{ssdil}. The SS dileptons can
also easily arise from $\tg\tq$ and $\tq\tq$ production. In Table \ref{tab:osss},
we list OS and SS dilepton rates in fb from SM sources along with SUSY points P1 and P2.
Here, we see that the SS/OS ratio is greatly enhanced when SUSY events are present.
Further, since LHC is a $pp$ collider, we expect an asymmetry in SS dilepton production
when squarks are produced at large rates. This is also seen in Table \ref{tab:osss},
where a measurable $++$ to $- -$ asymmetry is observed due to SUSY
contamination.
%
\begin{table}
\begin{center}
\begin{tabular}{lcccccc}
\hline
point & OS & ++ & $- -$ & SS/OS & ++/OS & $- -$/OS \\
\hline
SM    & 685  & 0   & 0 & 0.0 & 0.0 & 0.0 \\
SM+P1 & 1050 & 39 & 34 & 0.070 & 0.037 & 0.032 \\
SM+P2 & 1052 & 39 & 35 & 0.071 &0.037 & 0.033 \\
SM+P3 & 916 & 13 & 15 & 0.031 &0.014 & 0.016 \\
\hline
\end{tabular}
\caption{Cross sections in fb for OS and SS dilepton events from
SM sources and for SM plus points P1, P2 or P3, along with the SS to OS
dilepton ratio.
}
\label{tab:osss}
\end{center}
\end{table}
%

\section{Conclusions}
\label{sec:conclude}

One of the first orders of business for the LHC experiments will be to ``re-discover''
the Standard Model, {\it i.e.} to make sure that familiar processes like QCD
dijet production, $W+$jets production, $Z+$jets production and $\ttb$ production
all occur as expected. In fact, these processes will likely be used as 
{\it calibration} processes, to calibrate the detectors and to tune
Standard Model Monte Carlo programs. 
In this paper, we have emphasized that low mass SUSY may already make significant
contributions to these familiar SM processes, these SUSY contributions, if not taken into account, could cause detector
mis-calibration, or mis-tuning of SM Monte Carlo codes.

It will be imperative to check that the SM calibration reactions have {\it all}
of the properties expected from these reactions. We point out several ways in which
$Z+$jets, $W+$jets and $\ttb$ production can be checked for contamination 
from non-SM processes like low mass SUSY. 
\bi
\item In $Z+$jets events, even though
initially $\eslt$ will not be well-measured, there should still be an 
obvious excess of $\eslt$ from SUSY contamination.  
\item In $W+$jets events, we advocate plotting the $W/Z$ ratio versus
jet multiplicity. The SM prediction should grow
with jet multiplicity due to the presence of $\ttb$ contributions. 
However, SUSY contamination is likely to upset the expected ratios, 
especially at high jet multiplicity. 
\item In $\ttb$ events, we advocate comparing the shape
of the trijet distribution which reconstructs $m_t$ against SM expectations. 
In addition, the
dilepton-to-single-lepton event ratio can be sensitive to SUSY contamination, 
as can the presence of a high rate of same-sign dileptons. 
\ei

Only if all the properties from the SM processes accurately match the data can they be used with a high degree of confidence for calibrations.

\section*{Acknowledgments}
We thank Tom LeCompte for discussion.  This work was supported in part by the U.S.~Department of Energy under 
grant Nos. DE-FG02-97ER41022 and DE-FG02-95ER40896, the NSF under Grant No. PHY05-51164 and by the 
Wisconsin Alumni Research Foundation.  HB and VB thank the KITP Santa Barbara for hospitality.

%

%
\end{document}